# Highly scalable, wearable surface-enhanced Raman spectroscopy

*Limei Liu [†], Pablo Martinez Pancorbo[†], Ting-Hui Xiao [\*], Saya Noguchi, Machiko Marumi, Julia Gala de Pablo, Siddhant Karhadkar, Kotaro Hiramatsu, Hiroki Segawa, Tamitake Itoh, Junle Qu, Kuniharu Takei, Keisuke Goda [\*]*

[†]Co-first authors: equally relevant contributions independently on the order.

Dr. L. Liu, Dr. P. Martinez Pancorbo, Prof. T.H. Xiao, S. Noguchi, M. Marumi, Dr. J. Gala de Pablo, S. Karhadkar, Prof. K. Hiramatsu, Prof. K. Takei, Prof. K. Goda
Department of Chemistry, The University of Tokyo, Tokyo 113-0033, Japan

Dr. L. Liu, Prof. J. Qu
College of Physics and Optoelectronic Engineering, Shenzhen University, Shenzhen 518060, China

Prof. T.H. Xiao, Prof. K. Goda
Institute for Quantum Life Science, National Institute for Quantum and Radiological Science and Technology, Chiba 263-8555, Japan

Dr. H. Segawa
Third Department of Forensic Science, National Research Institute of Police Science, Chiba 277-0882, Japan

Dr. T. Itoh
Health and Medical Research Institute, National Institute of Advanced Industrial Science and Technology, Takamatsu 761-0395, Japan

Prof. K. Takei
Department of Physics and Electronics, Osaka Prefecture University, Osaka, 599-8531, Japan

Prof. K. Goda
Institute of Technological Sciences, Wuhan University, Hubei 430072, China

Prof. K. Goda
Department of Bioengineering, University of California, Los Angeles, California 90095, USA

\* E-mail: xiaoth@chem.s.u-tokyo.ac.jp; goda@chem.s.u-tokyo.ac.jp

**Abstract**

The last two decades have witnessed a dramatic growth of wearable sensor technology, mainly represented by flexible, stretchable, on-skin electronic sensors that provide rich information of the wearer's health conditions and surroundings. A recent breakthrough in the field is the development of wearable chemical sensors based on surface-enhanced Raman spectroscopy (SERS) that can detect molecular fingerprints universally, sensitively, and noninvasively. However, while their sensing properties are excellent, these sensors are not scalable for widespread use beyond small-scale human health monitoring due to their cumbersome fabrication process and limited multifunctional sensing capabilities. Here we demonstrate a highly scalable, wearable SERS sensor based on an easy-to-fabricate, low-cost, ultrathin, flexible, stretchable, adhesive, and bio-integratable gold nanomesh. It can be fabricated in any shape and worn on virtually any surface for label-free, large-scale, in-situ sensing of diverse analytes from low to high concentrations (10 – $10^6$ nM). To show the practical utility of the wearable SERS sensor, we test the sensor for the detection of sweat biomarkers, drugs of abuse, and microplastics. This wearable SERS sensor represents a significant step toward the generalizability and practicality of wearable sensing technology.

**1. Introduction**

The last two decades have witnessed a dramatic growth of wearable sensor technology, mainly represented by flexible, stretchable, on-skin electronic sensors that provide rich information of the wearer's health conditions and surroundings[1,2]. Current wearable sensors generally track the wearer's motions and vital signs, such as steps, blood pressure, blood oxygen saturation, respiratory rate, and heart rate, under everyday conditions[3], and have been recently upgraded to perform in-situ chemical sensing of the wearer's biofluids, such as sweat, saliva, tears, and urine, in a non-invasive manner[4]. Multiplexed chemical analysis of biomarkers in biofluids is essential for an accurate and comprehensive understanding of the wearer's complex physiological and pathological conditions[5]. Unfortunately, conventional wearable sensors are only sensitive to one type of chemical in an analyte at a time and need to be tailored for the target analyte due to their limited sensing mechanism (mainly originating from the electronic nature of the sensors). For this reason, they are incapable of distinguishing different chemicals simultaneously in a single measurement[6]. Alternatively, they can be designed to measure multiple chemical species, but these modifications lead to larger sizes, higher fabrication costs, more synthetic steps, and the need for prior knowledge of the target analyte[7].

Surface-enhanced Raman spectroscopy (SERS) has emerged as an attractive approach to next-generation wearable sensors in recent years due to enabling highly sensitive, multiplexed chemical sensing of complex analytes in a non-invasive and label-free manner without the need for prior knowledge of the analytes[1,8–12]. Very recently, two groups have achieved a significant milestone nearly at the same time[2,13]. Specifically, Wang et al. developed a wearable SERS sensor composed of a plasmonic metafilm formed by an ordered silver nanocube superlattice[13]. Koh et al. demonstrated a different wearable SERS sensor composed of plasmonic silver nanowires on top of a silk fibroin protein film[2]. While these SERS sensors show excellent molecular specificity and high detection sensitivity, their scalability remains challenging due to their intricate fabrication processes and limited multifunctional sensing capabilities. As they rely on either a low-throughput laser writing method or an unstable substrate transfer method for their fabrication, it is difficult to fabricate the sensors on a large scale at a low cost for diverse applications.

In this study, we report our demonstration of a highly scalable wearable SERS sensor based on an easy-to-fabricate, low-cost, ultrathin, flexible, stretchable, adhesive, and bio-integratable gold nanomesh (**Figure 1**a-b). This sensor was inspired by a recently reported inflammation-free, gas-permeable, lightweight, stretchable electronic sensor made of gold-coated biocompatible polyvinyl alcohol (PVA) nanofiber that is attachable to human skin or non-flat, non-rigid surfaces for long durations of time[14]. We exploited its undiscovered optical properties and optimized its dimensions to obtain SERS capabilities, while retaining its excellent features such as easy fabrication, cost effectiveness, thinness, high flexibility, high stretchability, high adhesivity, and high bio-integratability (**Figure 1**c-f). This two-dimensional sensor is moldable and can be fabricated in any shape and worn on virtually any surface for label-free, large-scale, in-situ sensing of diverse analytes from low to high concentrations ($10 – 10^6$ nM). To show the practical utility of the wearable SERS sensor, we tested the sensor for the detection of sweat biomarkers, drugs of abuse, and microplastics.

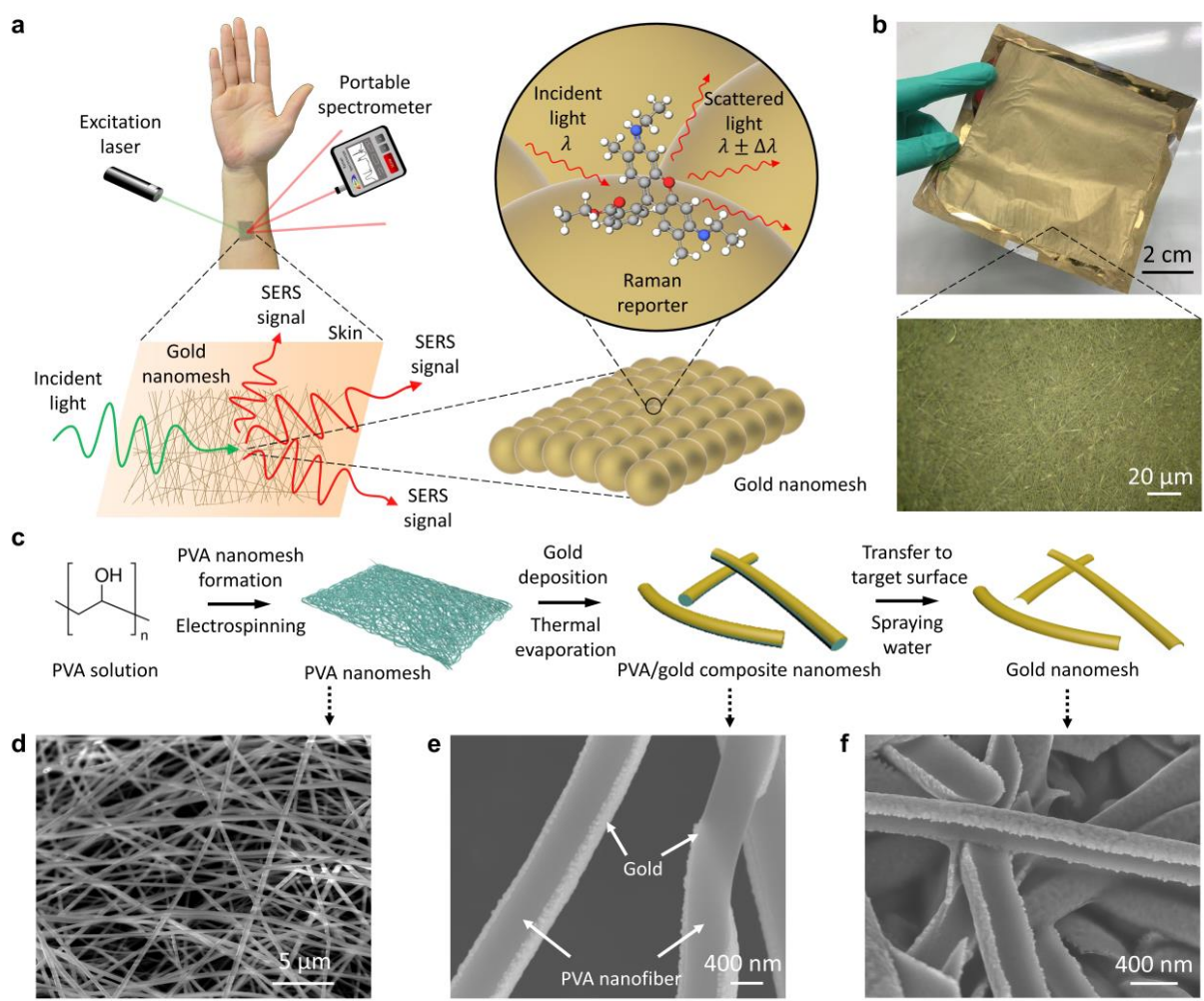

**Figure 1.** Concept, fabrication, design, and characterization of the wearable SERS sensor. a) Concept of wearable SERS on the skin. c) Picture of the fabricated gold nanomesh. The inset shows a 50x optical microscopy image of the gold nanomesh. c) Fabrication process and its corresponding SEM images of the gold nanomesh: d) PVA fiber nanomesh, e) gold-coated PVA fiber nanomesh, and f) gold nanomesh after removing the PVA fiber nanomesh.

## 2. Results and Discussion

As a proof-of-principle demonstration of the wearable SERS sensor's ability to detect molecules with high sensitivity, we used it to conduct SERS of rhodamine 6G (R6G) solutions on the sensor. With an integration time of 20 sec and an excitation power of 2 mW at an excitation wavelength of 785 nm, we first measured the Raman spectrum of an R6G solution at a concentration of 0.1 M on a silicon substrate as the ground truth. The characteristic Raman peaks of R6G were observed, as shown in **Figure 2**a. Then, by decreasing the laser excitation power to 0.2 mW and the R6G concentration to 100 nM, we observed the disappearance of the Raman spectrum on both the silicon and 150 nm-thick gold film substrates as expected. On the other hand, the Raman spectrum was visible on the wearable SERS sensor under the same conditions under the Raman signal enhancement. The SERS enhancement factor was found to be (2 mW / 0.2 mW) × (1 M / 100 nM) × (0.8 a.u. / 0.45 a.u.) = ~$10^8$ for R6G. It is important to mention that the enhanced Raman spectrum measured on the wearable SERS sensor agrees with the ground truth. **Figure 2**b,c shows that the lowest detectable concentration of R6G is about 10 nM (see **Movie S1** for our demonstration of Raman spectroscopy on the wearable SERS sensor).

Furthermore, to show the flexibility of the wearable SERS sensor as a SERS substrate, we conducted a crumpling test by adhering it to a hand glove and opening and closing the hand multiple times. As shown in **Figure 2**d,e, the Raman spectrum of R6G exhibited no appreciable change even after 1000 crumpling cycles. Likewise, we also performed a stretchability test on the wearable SERS sensor by adhering it to a 50% pre-stretched polydimethylsiloxane (PDMS) substrate and releasing/stretching it multiple times. As shown in **Figure 2**f,g the Raman spectrum of R6G exhibited no appreciable change even after 1000 stretching cycles (50% strain). These results firmly demonstrate the user friendliness and practical usability of the wearable SERS sensor.

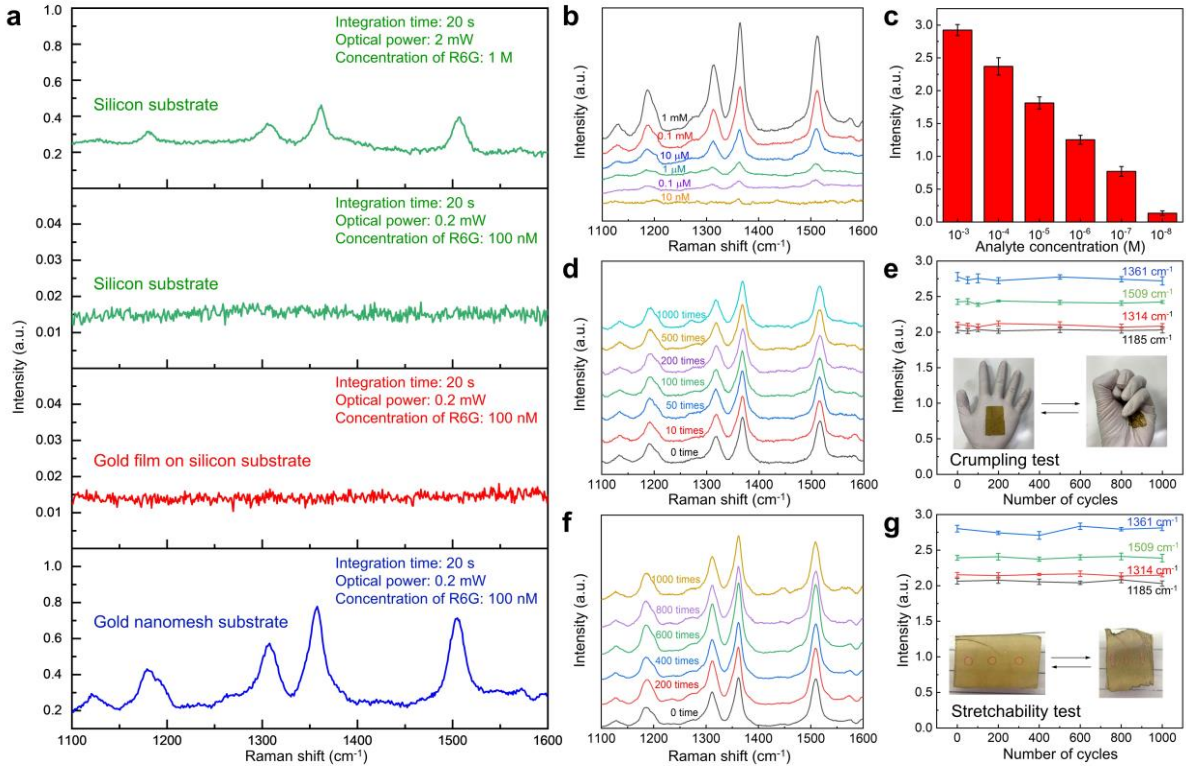

**Figure 2.** Basic SERS performance of the wearable SERS sensor. a) Raman spectra of R6G molecules on a silicon substrate, gold film substrate, and gold nanomesh substrate obtained under different measurement conditions (e.g., integration time, excitation power, R6G concentration) after baseline subtraction. Excitation wavelength: 785 nm. b) Raman spectra of R6G molecules on the wearable SERS sensor obtained at different R6G concentrations for an integration time of 20 sec with an excitation power of 0.2 mW at an excitation wavelength 785 nm after baseline subtraction. c) Intensities of the Raman peaks at a Raman shift of 1361 cm$^{-1}$ at different R6G concentrations, with standard deviations as error bars. The detection limit of the wearable SERS sensor for R6G molecules is about 10 nM. d) Raman spectra and e) characteristic Raman peaks of R6G molecules on the gold nanomesh adhered to a glove during the 1000-cycle crumpling test after baseline subtraction, with standard deviations as error bars. The inset shows pictures of the wearable SERS sensor on the glove when the hand was opened and closed. f) Raman spectra and g) characteristic Raman peaks of R6G molecules on the gold nanomesh adhered to a pre-stretched PDMS substrate during the 1000-cycle stretchability test, with standard deviations as error bars. The inset shows pictures of the wearable SERS sensor on the PDMS substrate when the PDMS substrate was stretched and released. The error bars in panels c, e, and g are the standard deviations calculated from five different samples.

To show the diverse utility of the wearable SERS sensor for SERS applications, we demonstrated high intrinsic adhesivity without using any glues by placing it on the surfaces of various objects. **Figure 3**a-c shows the wearable SERS sensor adhered to a human wrist for the detection of sweat biomarkers. The pictures show that it was firmly attached to the skin even under normal flexion and extension conditions, demonstrating high flexibility, stretchability, adhesivity, and bio-integrability. No obvious side effect on the human skin was observed by virtue of the inflammation-free and gas permeable properties of the gold nanomesh[14]. **Figure 3**d,e shows that the wearable SERS sensor adhered to a human cheek and a contact lens for the detection of biomarkers in tears[15,16]. Similarly, **Figure 3**f shows the wearable SERS sensor adhered to the surface of a face mask for the detection of breath and saliva biomarkers highly relevant to COVID-19 and other respiratory and oral diseases[17–21]. **Figure 3**g-i shows the wearable SERS sensor adhered to metallic and plastic surfaces, such as an elevator control panel, a door handle, a doorknob, and a computer keyboard, exhibiting its potential to serve as an environmental monitoring and infection surveillance tool in smart cities[22]. Finally, **Figure 3**k,i shows the wearable SERS sensor adhered to an apple and a leaf for food safety applications[1,23,24]. To show the practical utility of the wearable SERS sensor, we demonstrate three applications, namely sweat biomarker detection, drug identification, and microplastic detection.

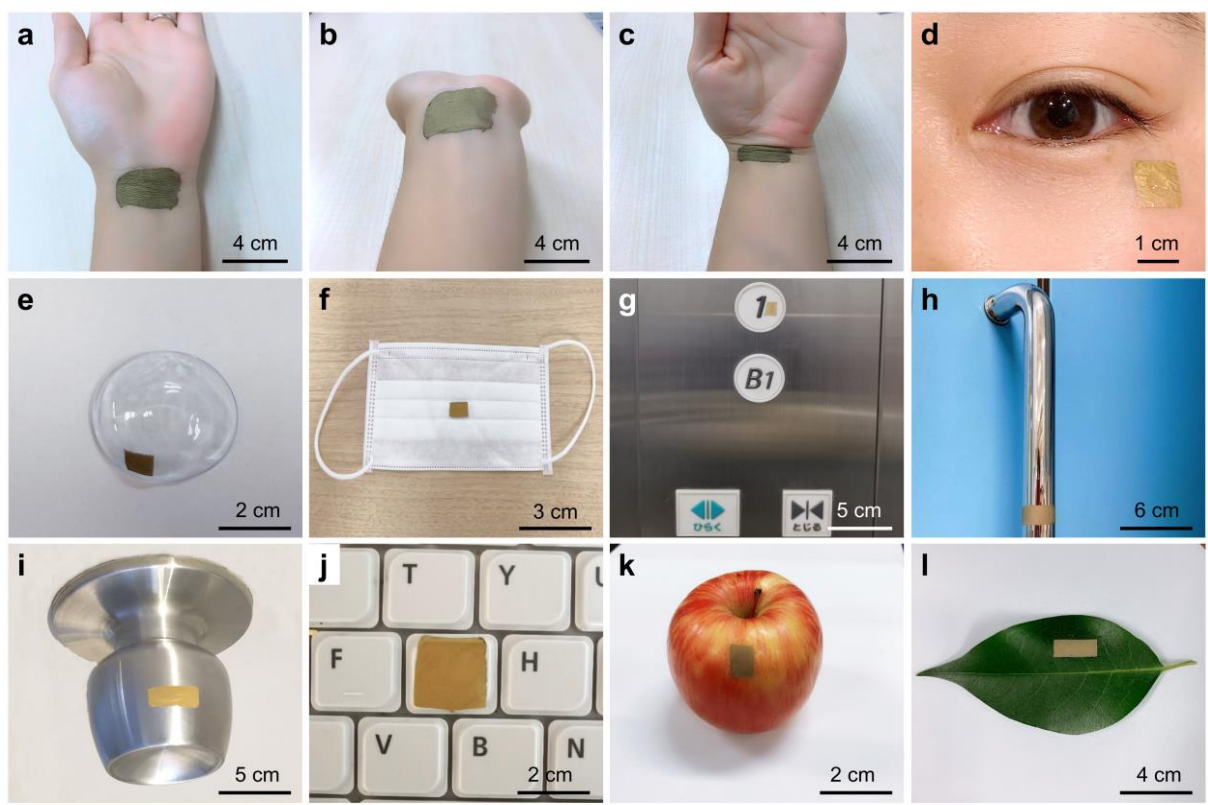

**Figure 3.** Pictures of the wearable SERS sensor to demonstrate its wearability for diverse applications. a) Wearable SERS sensor adhered to a human wrist, followed by undergoing

wrist flexion and extension (b-c). d-e) Wearable SERS sensor adhered to a human cheek and contact lens to detect biomarkers in tears. f-j) Wearable SERS sensor adhered to a face mask, elevator control panel, door handle, doorknob, and computer keyboard for environmental monitoring and infection surveillance, respectively. k-i) Wearable SERS sensor adhered to an apple and leaf for food safety applications.

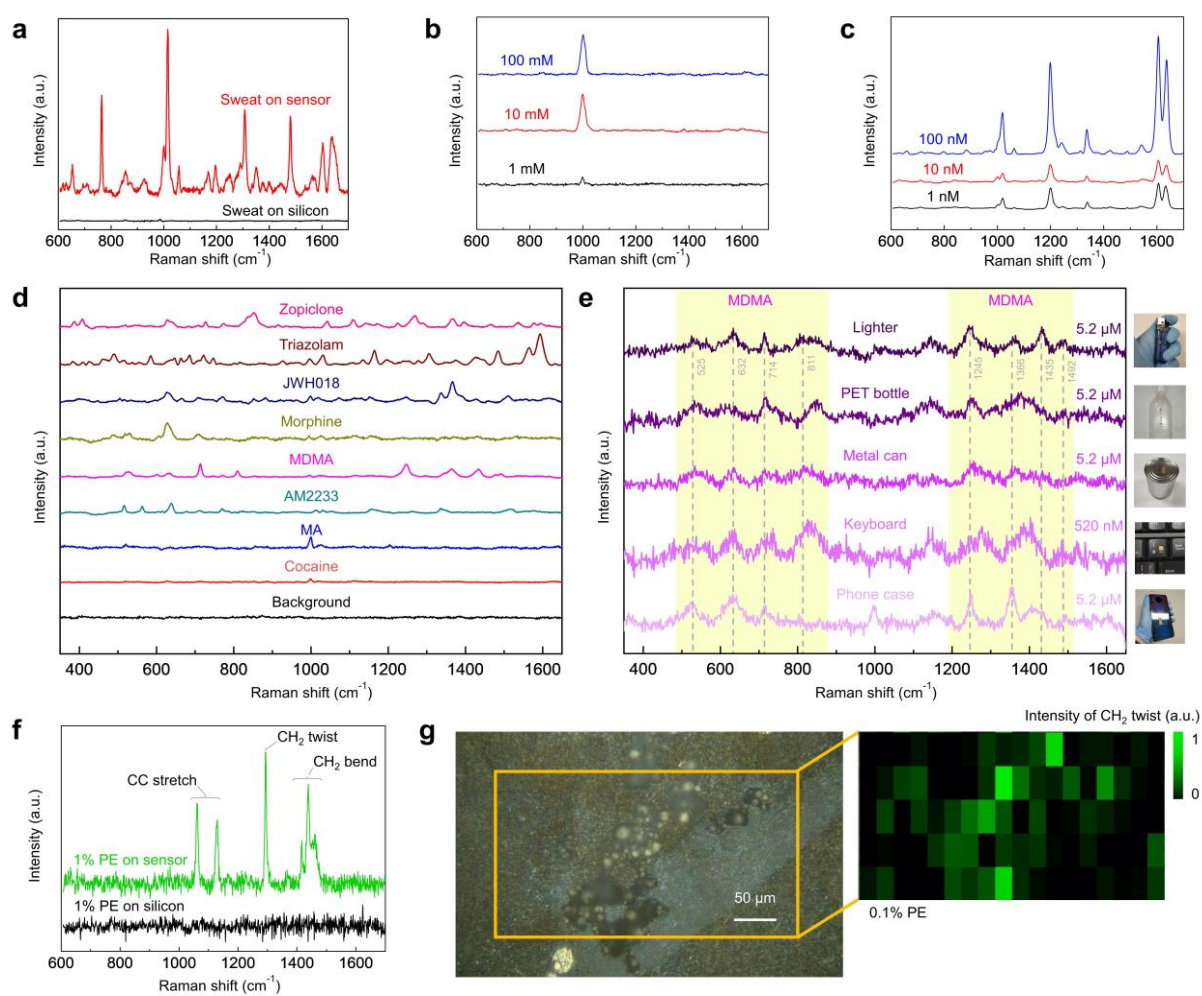

**Figure 4.** Potential applications of the wearable SERS sensor. a-c) Raman spectra of human sweat, aqueous urea solution, and ascorbic acid aqueous solution, respectively, obtained on the wearable SERS sensor for an integration time of 20 sec with an excitation power of 5 mW at an excitation wavelength of 785 nm after baseline subtraction. d) Raman spectra of various drugs of abuse in methanol solution obtained on the wearable SERS sensor for an integration time of 1 sec with an excitation power of 10 mW at an excitation wavelength 785 nm after baseline subtraction. e) Raman spectra of MDMA aqueous solution for an integration time of 3 sec at an excitation power of 10 mW with 8 acquisitions for 5.2 µM and for an integration time of 10 s with 4 acquisitions for 520 nM on the wearable SERS sensor attached to the surfaces of various daily life products, including a plastic lighter, silicon phone case, PET plastic bottle, metal can, and ABS plastic computer keyboard. f) Raman spectra of polyethylene microbeads obtained on the wearable SERS sensor and a silicon substrate for an integration time of 20 sec with an excitation power of 5 mW at an excitation wavelength of 785 nm after baseline subtraction. g) Raman map (at a Raman shift from 1286 cm$^{-1}$ to 1302 cm$^{-1}$) of polyethylene microbeads obtained on the wearable SERS sensor dispersed in deionized water at a concentration of 0.1% for an integration time of 10 sec with an excitation power of 0.5 mW at an excitation wavelength of 785 nm after baseline subtraction.

To demonstrate the capability of the wearable SERS sensor for health monitoring as a potential application, we measured human sweat from a human subject as well as two human sweat biomarkers: urea and ascorbic acid. The human sweat sample contains various substances such as sodium chloride, ascorbic acid and urea. **Figure 4**a presents the Raman spectra of 2-µL human sweat obtained from the participant on the wearable SERS sensor. In contrast, no observable Raman peaks from the same human sweat sample were obtained on a silicon wafer. Moreover, we used the wearable SERS sensor to detect urea dissolved in water at various concentrations of 1 mM, 10 mM, 100 mM, and 1 M from 2-µL drops. **Figure 4**b shows the measured Raman spectra of the urea solutions with the characteristic Raman peak at 1,003 cm$^{-1}$ after baseline corrections. Likewise, we obtained Raman spectra of ascorbic acid at minute concentrations of 1 nM, 10 nM, 100 nM, and 1,000 nM from 2-µL drops with characteristic Raman peaks at 1,020 cm$^{-1}$, 1,200 cm$^{-1}$, 1,337 cm$^{-1}$, 1,606 cm$^{-1}$ and 1,636 cm$^{-1}$, after baseline corrections (**Figure 4**c). These results illustrate that the wearable SERS sensor can detect urea and ascorbic acid with low concentrations by comparing with the human sweat spectra from **Figure 4**a. These ranges cover the typical concentrations of urea and ascorbic acid on human skin after sweat evaporation, indicating the practical utility of the wearable SERS sensor for sweat analysis on the human skin.

Next, to demonstrate the capability of the wearable SERS sensor for forensic science as another potential application, we performed SERS of some of the widely abused drugs of abuse in the world [25], including methamphetamine (MA), *3,4*-Methylenedioxy methamphetamine (MDMA), cocaine, (*2*-iodophenyl)(*1*-((*1*-methylpiperidin-*2*-yl)methyl)-*1H*-indol-*3*-yl)methanone (AM2233), morphine, *1*-pentyl-*3*-(*1*-naphthoyl)indole (JWH-018), triazolam, and zopiclone. They were dissolved in methanol and measured on the wearable SERS sensor at concentrations of 0.67 mM for methamphetamine, 0.33 mM for cocaine, 0.52 mM for MDMA, 0.22 mM for AM-2233, 0.35 mM for morphine, 0.29 mM for JWH-018, 0.29 mM for triazolam, and 0.26 mM for zopiclone. Their Raman spectra with characteristic peaks obtained after baseline subtraction are shown in **Figure 4**d. No Raman signal was obtained at these concentrations on a silicon substrate, which demonstrates the effectiveness of SERS in the detection of drugs of abuse. Furthermore, MDMA dissolved in water was detected at low concentrations of 520 nM and 5.2 µM on the wearable SERS sensor attached to the surfaces of various daily life products, including a plastic lighter, silicon phone case, PET plastic bottle, metal can (aluminum), and ABS plastic computer keyboard as shown in **Figure 4**e, which confirms the surface-agnostic capability of the wearable SERS sensor. Here, the main Raman peaks of MDMA are at 525, 632, 714, 811, 1,245, 1,366, 1,435 and 1,492 cm$^{-1}$ while the other Raman peaks mainly come from the surfaces of the various daily life products where the sensor was attached.

Finally, to demonstrate the capability of the wearable SERS sensor for environmental monitoring, we used it to identify microplastics in water with different concentrations. Specifically, microplastic pollution is known to spread extensively across the oceans on Earth, affecting many ecosystems[26,27]. The most abundant microplastic species found floating in the ocean is polyethylene, constituting approximately 54.5% of all floating microplastic pollution. **Figure 4**f shows the Raman spectrum of polyethylene microbeads (**Figure S1**a,b) in distilled water on a silicon substrate and the wearable SERS sensor. Here, the polyethelene microbeads were aggregated on the liquid surface due to their high hydrophobicity, but some of them remained dispersed in the liquid at a lower concentration. The highest Raman peak that appeared at 1,296 cm$^{-1}$ was used for Raman mapping. **Figure 4**g shows an optical image and Raman map of the wearable SERS sensor attached to a silicon wafer with polyethylene microbeads of density 950 g/L$^3$ at a concentration of 0.1%. It is evident from the Raman map that the wearable SERS sensor is robust for the detection of microplastics. Even if the nanomesh structure of the sensor is partially damaged under rough conditions, it can provide reproducible SERS signal enhancements for polyethylene microbeads. In addition, there is no significant difference in measured Raman spectra of dry and wet microplastic samples as long as the microplastics remain in close proximity to the surface of the substrate at the time of SERS measurements. In comparison, a Raman mapping on a silicon wafer with identical experimental conditions is shown in **Figure S1**c, which exhibits a much lower Raman intensity. Moreover, **Figure S1**d,e shows an approach as a potential practical application to evaluating water pollution by attaching the wearable SERS sensor to a water-immersible surface and submerging it for 10 sec.

## 3. Conclusions

In summary, we demonstrated highly scalable, wearable SERS by using an easy-to-fabricate, low-cost, ultrathin, flexible, stretchable, adhesive, and bio-integratable gold nanomesh that enabled the detection and identification of diverse analytes at low concentrations (10 nM for R6G and ascorbic acid solutions) on virtually all arbitrary surfaces. The wearable SERS sensor exhibited excellent practical utility for various types of applications, including the detection of sweat biomarkers, drugs of abuse, and microplastics that were explored in this study. This wearable SERS sensor represents a significant step toward the generalizability and practicality of wearable sensing technology.

There is one limitation to the demonstrated wearable SERS sensor. It requires an external light source to excite Raman scattering of analytes and an external spectrometer to collect scattered light. To make it an independent device, one possible solution is to further integrate a semiconductor nanolaser and a nanospectrometer into the wearable SERS sensor by direct bonding. With this improvement, our wearable SERS sensor is expected to be highly valuable as an essential ingredient of wearable sensing technology.

## 4. Experimental Section

*Chemicals:* R6G, PVA powders, and zopiclone were purchased from Sigma Aldrich. The polyethylene microbeads used in this study are the Polyethylene Nanospheres 0.95g/cc 200-9900 nm - 1g (~ 1-10 µm size) from Cospheric LLC. Triazolam, urea (Wako 1st Grade), and ascorbic acid (0.1 M) were purchased from FUJIFILM Wako Pure Chemical Corporation. Deionized water was collected at 15.0 MΩ and 14.6 °C from Milli-Q (Merck Millipore) using vent filters MPK01 and Q-POD1.

*Fabrication of the wearable SERS sensor:* The fabrication process of the wearable SERS sensor and its integration into a target surface are described as follows and shown in **Figure 1**c. First, a PVA aqueous solution was formed by dissolving 8 wt% PVA into deionized water and stirring at 80 °C for 12 hours. Then, PVA nanofibers with a diameter of ~500 nm were prepared by electrospinning an 8 wt% PVA aqueous solution and intertwined to form a mesh-like sheet by electrospinning at 25 kV for 40 min by using microfluidic pumps at 1 mL h-1 flow rate. Second, a 150-nm-thick gold layer was thermally deposited on the surface of the PVA nanofibers at a gold flow rate of 0.1 nm/sec at a high vacuum ($10^{-6}$ torr) to produce a gold nanomesh. Finally, water was sprayed on a target surface, such as the human skin, to attach the gold nanomesh on the surface, followed by spraying water again, but on the gold nanomesh to dissolve the PVA nanofiber, resulting in a pure gold nanomesh structure without PVA. To transfer the gold nanomesh to polyethylene film (7 µm thick), skin or other various substrates (e.g., a leaf, fruits, glove, or face mask), the gold nanomesh was attached to the surface, followed by spraying water on them to rinse away PVA nanofibers. The attached gold nanomesh was then dried at room temperature.

*Raman measurements:* Raman spectra from all the samples except the drugs were obtained using a RM 2000 microscopic confocal Raman spectrometer (InVia, Renishaw PLC, England) excited by a 785-nm-wavelength continuous-wave laser with a maximum power of 500 mW. The drugs of abuse were evaluated using Laser Raman Microscopy RAMANforce from Nanophoton Corp. The portable spectrometer used in Movie S1 is the NanoRam®

Handheld Raman Spectrometer from B&W Tek. Scanning electron microscope (SEM) images of the wearable SERS sensor fabrication and microplastic particles were taken from a JSM-7600F FESEM microscope. The wearable SERS sensor attached to the glove was worn on a hand for the flexibility test. After opening and closing the hand multiple times, the Raman spectra of R6G on the wearable SERS sensor with the glove were obtained. The stretchability of the wearable SERS sensor was examined on a PDMS substrate, which was pre-stretched by 50%. When releasing the PDMS substrate, the gold nanomesh formed wrinkled structures and did not break even if it was fully released. The Raman spectra of R6G on the gold nanomesh with the PDMS substrate were obtained at different compression strains in compression cycles (up to 1000 times). The Raman measurements in **Figure 2** were performed under 50x magnification for performance characterizations, while those in **Figure 4** were conducted under 20x magnification for practical applications.

*Material characterization*: **Figure 1**f-h shows the SEM images corresponding to the PVA nanofibers, PVA/gold nanofibers, and semi-hollow gold nanofiber, respectively. We optimized the diameter of the PVA nanofiber (~500 nm) and the thickness of the gold layer (150 nm) to maximize the effect of localized surface plasmon resonance (LSPR) on the surface of the semi-hollow gold nanofiber. **Figure 1**b shows that the wearable SERS sensor can be produced in large areas, while its inset shows the gold nanomesh under an optical microscope using a 50x magnification lens.

*Human sweat samples:* The human sweat samples were collected with a micropipette from exercise-induced sweat on the clean skin of a voluntary human subject. The SERS test was approved by the Research Ethics Committee in the Graduate School of Science at the University of Tokyo (approval number: 20-352. The human subject expressed informed consent prior to the sample collection.

**Supporting information**

Supporting information is available from the Wiley Online Library or from the author.

**Data availability**

The source data supporting the findings of this study are available from the corresponding authors upon reasonable request.

**Acknowledgements**

This research was supported by MEXT Quantum Leap Flagship Program (JPMXS0120330644), JSPS KAKENHI (JP18K13798, JP20K14785), JSPS Core-to-Core Program, KISTEC, Murata Science Foundation, White Rock Foundation, University of Tokyo GAP Fund, and UTokyo IPC. PMP acknowledges financial support from the Grant-in-Aid for JSPS Fellows (P20789). JGP acknowledges financial support from the JSPS Grant-in-Aid for Young Scientists (20K15227) and the Grant-in-Aid for JSPS Fellows (19F19805).

**Author contributions**

LL and PMP performed the experiments and analyzed the experimental data. LL, PMP, THX, and KG drafted the paper. All co-authors contributed to modifications of the paper. LL, PMP, SN, and MM synthesized the wearable SERS sensors and optimized them. LL measured the Raman signals of R6G. LL and THX designed and performed the stretchability and crumping tests. LL, PMP and THX took photographs of the wearable SERS sensor on various surfaces. PMP designed and performed the experiments with human sweat. PMP and HS designed and performed the experiments with drugs using the wearable SERS sensor synthesized by PMP. PMP, JGP, and THX designed and performed the measurements with microplastics using the wearable SERS sensor synthesized by PMP. PMP designed and performed the water immersion test. PMP and SK obtained the SEM images. KH, HS, TI, and KT provided help with the manuscript and data interpretation. THX, JQ, and KG supervised the research project.

**Competing interests**

KG is a shareholder of CYBO and Cupido. THX, LL, KH, and KG hold a pending patent for the gold nanomesh sensor.